\documentclass[%
aps,
pre,
reprint,
amsmath,
amssymb,
superscriptaddress]{revtex4-2}
\usepackage{graphicx}
\usepackage{xcolor}
\usepackage{bm}
\usepackage{siunitx}
\usepackage[%
hidelinks,
colorlinks=true,
citecolor=blue,
urlcolor=blue,
linkcolor=blue
]{hyperref}

\definecolor{red}{HTML}{C74431}

\definecolor{green}{HTML}{3C9455}

\begin{document}
\author{Roland Wiese}
\email{wiese@itp.uni-leipzig.de}
\author{Klaus Kroy}
\affiliation{Institute for Theoretical Physics, Leipzig University, 04103 Leipzig, Germany}
\author{Viktor Holubec}
\email{viktor.holubec@mff.cuni.cz}
\affiliation{Faculty of Mathematics and Physics, Charles University, CZ-180 00 Prague, Czech Republic}
\date{\today}

\title{Active Ornstein-Uhlenbeck Model for Bacterial Heat Engines}
\begin{abstract}
    We use Brownian dynamics simulations to study a model of a cyclic bacterial heat engine based on a harmonically confined colloidal probe particle in a bath formed by active Brownian particles. For intermediate activities, active noise experienced by large enough probes becomes Gaussian with exponential autocorrelation function.
    We show that, in this experimentally pertinent regime, the probability densities for stochastic work, heat, and efficiency are well represented by those of a single active Ornstein-Uhlenbeck particle (AOUP), effectively representing the whole many-body setup.
    Due to the probe's fast relaxation in the potential, in typical experimental implementations, good agreement can prevail even when the autocorrelation function of the active noise develops non-exponential tails.
    Our results show that the AOUP provides a convenient and accurate, analytically tractable effective model to mimic and analyze experimental bacterial heat engines, especially when operating with comparatively large probes and stiff traps. 
\end{abstract}
\maketitle
\section{Introduction}
Recent developments in micro-manipulation techniques now allow the construction of tiny heat engines using a single Brownian particle \cite{blickle2012realization,martinez2017colloidal-heat-engines-review}, or even electrons in a quantum dot~\cite{Josefsson2018}, as the working medium.
Thermodynamic variables characterizing these devices fluctuate strongly during a single operation cycle~\cite{holubec2021fluctuations-in-heat-enginges}. 
The framework of stochastic thermodynamics~\cite{sekimoto2010stochastic-energetics, seifert2012stochastic-thermodynamics} extends the macroscopic notions of work and heat to describe these fluctuating engines and their efficiencies.
The classical results are recovered upon averaging~\cite{callen1985thermodynamics}. 
A recent experiment studying a colloidal heat engine operating in a bacterial bath~\cite{krishnamurthy2016micrometre} led to the generalization of this framework to encompass Brownian heat engines operating in active baths~\cite{marathe2019stochastic-work-extraction, holubec2020active-brownian-heat-engines, fodor2021active-engines, fodor2022irreversibility}.

When the probe-bath interaction can be captured by an effective temperature, the heat engine with an active bath obeys the same (average) thermodynamics and achieves the same performance as with a fictitious equilibrium bath at the same temperature~\cite{holubec2020active-brownian-heat-engines, partI}.
This mapping thus allows one to derive bounds on the performance of an active heat engine from the second law of thermodynamics.
Furthermore, it provides a route how to represent the thermodynamics of complicated non-equilibrium heat engines by much simpler, effective models. 

In this work, we study how far such a reduced model for a prototypical active heat engine reproduces not only the (average) thermodynamic variables but also their stochastic fluctuations.
In contrast to previous computer simulations, where the active bath was assumed to provide an exponentially correlated noise~\cite{Saha2018, Kwon2024}, we go one step further and consider a cyclic heat engine based on a harmonically confined colloidal particle interacting with an explicitly modeled active bath composed of interacting active Brownian particles. 
The latter serves to represent individual bacteria in the bacterial heat engine of Ref.~\cite{krishnamurthy2016micrometre}, which motivated our model, or similar entities in alternative designs.
The (average) thermodynamic performance of such active heat engines is analyzed in depth in a companion paper~\cite{partI}.

A general theoretical approach to developing low-dimensional coarse-grained models for complicated many-body systems is to integrate out degrees of freedom that are not of interest in a systematic bottom-up approach~\cite{SCHILLING20221}.
This procedure requires a faithful mathematical description of the complete experimental system to which the coarse-graining shall be applied, which might not be available in the case of a bacterial heat bath. 
To overcome the ensuing technical difficulties, theoretical work often resorts to the (questionable) limit of weak bath-probe coupling~\cite{Steffenoni2016,martin2018extracting} and yields viscoelastic probe dynamics with a friction kernel and a (generally non-Gaussian~\cite{roldan2024shot-noise-active-bath}) noise with multiple time-scales. It leaves the practitioner somewhat puzzled with regard to its experimental implications. 
The following top-down approach is intended to provide a remedy in this regard.

As our main numerical result, we find that for a wide range of experimentally relevant parameter values, the stochastic thermodynamics of the probe is surprisingly well described by a harmonically confined active Ornstein-Uhlenbeck particle (AOUP)~\cite{fodor2016how-far} with exponentially correlated Gaussian noise and without necessitating a friction renormalization.
The coarse-grained effective model reproduces quantitatively the work, heat, and efficiency distributions of the simulated many-body active heat engine, even when the noise autocorrelation function is non-exponential. 
The AOUP model is linear, and thus, besides providing a significant speed-up of computer simulations, it is also amenable to analytical treatment by techniques described in Ref.~\cite{holubec2021fluctuations-in-heat-enginges}, providing explicit exact results for simple driving protocols. 
It thus becomes a very convenient tool for fitting and interpreting experimental data, which is particularly useful for measuring probability densities or large deviation functions, for which getting good statistics is a challenge.

The manuscript is structured as follows. In Sec.~\ref{sec:model}, the full many-body setup of a colloidal heat engine operating in an active bath is introduced. 
The statistics of the colloid's position and the noise autocorrelation function are discussed in Sec.~\ref{sec:coarse}. 
As our main result, we show in Sec.~\ref{sec:fluctuations} how to use the AOUP model to reproduce the stochastic work, heat, and efficiency.
We conclude in Sec.~\ref{sec:conclusion}.

\section{Model of a bacterial heat engine}
\label{sec:model}
\begin{figure}[t]
    \centering
    \includegraphics[width=\linewidth]{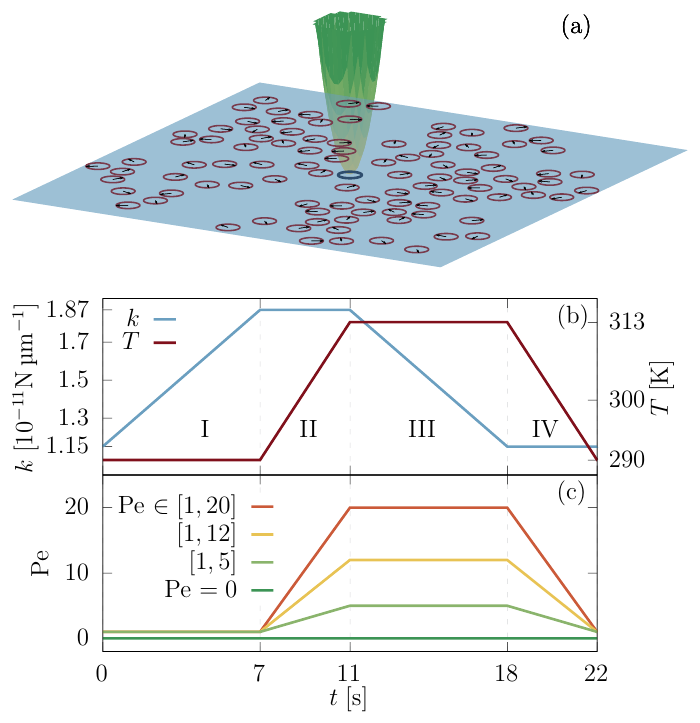}
    \caption{
    Active heat engine, numerical setup, and protocols (graphics adapted from Ref.~\cite{partI}).
    (a) Snapshot of the simulated two-dimensional setup, showing the passive tracer (blue disk), which is trapped by a parabolic potential (green) and interacts with a bath of active particles (red disks) in a solvent (blue).
    (b) Cycle protocols of the trap stiffness $k(t)$ and solvent temperature $T(t)$ of the colloidal Stirling engine~\cite{blickle2012realization}.
    (c) The activity of the bath particles is governed by a time-dependent P{\'e}clet number $\mathrm{Pe}(t)$ that follows the solvent temperature.
    }
    \label{fig:setup-protocol}
\end{figure}
We study a colloidal probe particle trapped in a harmonic potential $U(\bm r) = k\bm r^2/2$ of stiffness $k$ and diffusing in a suspension of ABPs, as sketched in Fig.~\ref{fig:setup-protocol}(a).
The dynamics of the probe's position vector obeys the overdamped Langevin equation
\begin{equation}
    \dot{\bm r} = \mu_0\bm F - \mu_0 k\bm r + \sqrt{2D_t^0}\bm \xi\,.
\label{eq:eom-probe}
\end{equation}
As the only modification from the companion paper~\cite{partI}, the probe's diameter $d_0$ and accordingly its mobility $\mu_0 = \mu d / d_0$ and translational diffusivity $D_t^0 = k_BT\mu_0 = D_t d/d_0$ are allowed to differ from \textcolor{black}{the diameter $d$, mobility $\mu$, and translational diffusivity $D_t$} of the $N$ bath ABPs.
The dynamics of the latter is governed by the overdamped Langevin equations (for $i=1,\ldots,N$)
\begin{align}
    \begin{split}
        \dot{\bm r}_i &= \mu\bm F_i + v_a\bm n_i + \sqrt{2D_t}\bm \xi_i\,,\\
        \dot{\theta}_i &= \sqrt{2D_r}\nu_i\,.
    \end{split}
    \label{eq:eom-abp}
\end{align}
The mutually independent unbiased normalized Gaussian white noises $\bm\xi$, $\bm\xi_i$, and $\nu_i$ model the thermal noise of the solvent at temperature $T$, in which the colloid and the ABPs are suspended. 
Only the latter self-propel at a constant speed $v_a$ along their orientations $\bm n_i=(\cos\theta_i,\,\sin\theta_i)$, while their orientation angles $\theta_i$ diffuse with rotational diffusivity $D_r$. 
As in the experiment, their motion is assumed to be unaffected by the trap potential. 
\textcolor{black}{Their activity is thus fully characterized by the non-dimensional P{\'e}clet number $\mathrm{Pe}=v_a/(d D_r)$.}

The particles only interact when closer than $r_{ij} = |\bm r_i - \bm r_j| < 2^{1/3}d_{\rm eff}$.
Here, and in Eq.~\eqref{eq:pair-pot}, $i=0$ is understood to refer to the probe. For mutual interactions between ABPs, the effective diameter $d_{\rm eff}$ is given by the bath particle radius $d$, and for interactions with the probe by $d_{\rm eff} = (d_0 + d)/2$. The repulsive interparticle forces $\bm F$ and $\bm F_i$ derive from an isotropic soft pair potential
\begin{equation}
    V(r_{ij}) = 4\epsilon\left[ \left(\frac{d_{\rm eff}}{r_{ij}}\right)^6 - \left(\frac{d_{\rm eff}}{r_{ij}}\right)^3 + \frac{1}{4}\right]
    \label{eq:pair-pot}
\end{equation}
with energy scale $\epsilon\gg k_BT$.

The described setup was designed to model the colloidal heat engine in a bacterial bath previously studied in experiments~\cite{krishnamurthy2016micrometre}.
As in Ref.~\cite{partI}, we therefore choose model parameters that mimic the experimental conditions. 
For the micron-sized active particles ($d=\SI{1}{\micro\metre}$), the Stokes mobility is $\mu=\SI{5e7}{\second\per\kilogram}$ in water at temperature $\SI{290}{\kelvin}$, and a realistic rotational diffusivity is $D_r=\SI{1}{\hertz}$, which are the values used throughtout our study. 

The dynamical equations are solved (iteratively) using Brownian dynamics simulations with time step $\mathrm{d}t = \SI{20}{\micro\second}$ inside a square box of side length $L$ with periodic boundary conditions.
We adjust the box size to maintain a moderate bath packing fraction \textcolor{black}{$\phi \equiv \pi (Nd^2 + d_0^2) /4L^2=0.2$} with particle number $N=10^2$ for ``small'' probes ($d_0<\SI{5}{\micro\metre}$) and $N=500$ for ``larger'' ones ($d_0\geq\SI{5}{\micro\metre}$).

Thermodynamic control parameters are the stiffness of the trap $k$ for the probe and the swim speed $v_a$ of the ABPs.
The probe particle is subjected to a cyclic driving protocol of period $t_p=\SI{22}{\second}$, mimicking the Stirling cycle~\cite{blickle2012realization, krishnamurthy2016micrometre, partI}.
It consists of piecewise constant and linear profiles in the trap stiffness $k(t)$ and solvent temperature $T(t)$, which are modulated between $k_{\min}=\SI{1.15e-11}{\newton\per\micro\metre}$, $k_{\max}=\SI{1.87e-11}{\newton\per\micro\metre}$ and $T_c=\SI{290}{\kelvin}$, $T_h=\SI{313}{\kelvin}$, as depicted in Fig.~\ref{fig:setup-protocol}(b).
The durations of isothermal and isochoric processes are set to $\SI{7}{\second}$ and $\SI{4}{\second}$, respectively, as in the experiment~\cite{krishnamurthy2016micrometre}, and also the ABPs' activity $\mathrm{Pe}(t)$ is assumed to follow the solvent temperature $T(t)$ in the range $\mathrm{Pe}\in[1,10^2]$, as shown in Fig.~\ref{fig:setup-protocol}(c). 
Energy is measured in units of $k_BT_c=\SI{4.002}{\newton\micro\metre}$ and observables are saved at a rate of $\Delta t=\SI{2}{\milli\second}$, as in the experiments~\cite{krishnamurthy2016micrometre}.

\section{Statistics of the tracer particle}
\label{sec:coarse}
In this section, we find a range of relative probe sizes and bath activities for which the active noise $\bm F$ in Eq.~\eqref{eq:eom-probe} is Gaussian and exponentially correlated, which are the conditions for coarse-graining the heat engine to the AOUP model.
The dynamical equation~\eqref{eq:eom-probe} for the probe's position is linear; hence, the noise is Gaussian when the position is Gaussian distributed. 
In Sec.~\ref{sec:posPDF}, we, therefore, assess the non-Gaussianity of the probe position distribution in a stationary trap potential by means of its fourth cumulant and diffusive rescaling. In Sec.~\ref{sec:ACFs}, we study the active noise's autocorrelation function (ACF).
It is found that, to a good approximation, the relevant properties of the active bath are independent of the probe's dynamics, the trap stiffness, and the background noise intensity.
Therefore, the depicted measurements are limited to constant driving parameters $k=k_{\min}$ and $T=T_c$, corresponding to their minima in the cyclic protocol.

\subsection{Position distribution of the probe}
\label{sec:posPDF}
We consider first the effect of the relative probe size and the P\'eclet number of the ABPs in the bath on the probe's position distribution. 
Since the intensity of the individual particle collisions has a finite variance, the central limit theorem implies that the probability density of the active noise intensity experienced by the probe should become Gaussian with increasing relative probe size $d_0/d$ due to the increasing collision frequency.
Due to the radial symmetry of the potential, the probability densities for $x$ and $y$ components of the position vector $\bm r$ are the same.

\begin{figure}[t]
    \centering
    \includegraphics[width=\linewidth]{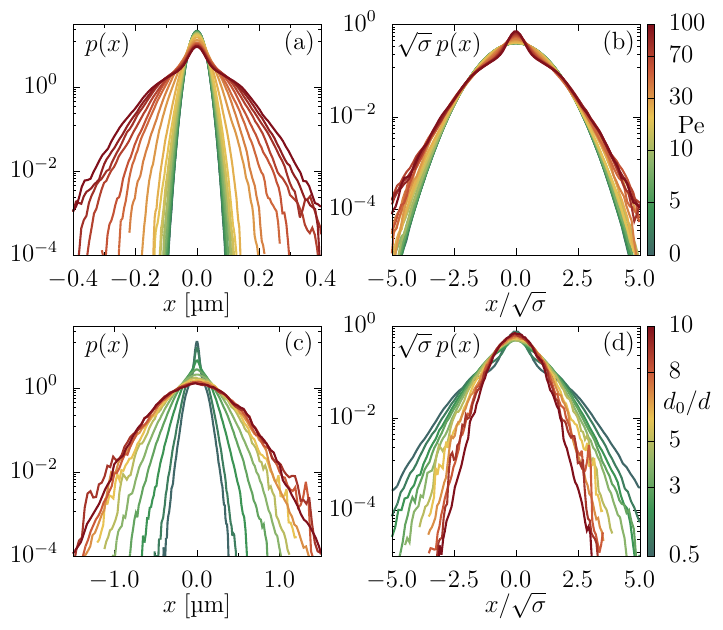}
    \caption{
    Position distributions $p(x)$ for the $x$-coordinate of the trapped probe at $k_{\min}$ and $T_c$:
    (a), (b) for small probe size $d_0=d=\SI{1}{\micro\metre}$ and varying activity $\mathrm{Pe}$ (color code), they develop heavy tails for higher activities compared to the normal distribution at $\mathrm{Pe}=0$;
    (c), (d) for large constant $\mathrm{Pe}=100$ and increasing probe diameter $d_0$ (color code) they approach a Gaussian shape;
    (b), (d) rescaling the histograms by their standard deviation $\sqrt{\sigma}$ makes them collapse in case they are Gaussian~ \cite{leptos2009dynamics-of-enhanced-tracer-diffusion,thiffeault2015distribution}.
    }
    \label{fig:position-distr}
\end{figure}

In Fig.~\ref{fig:position-distr}, we show the distributions $p(x)$ for the $x$ coordinate of the probe. 
The color codes for the activity in the top panels (a) and (b) and the probe diameter in the bottom panels (c) and (d). 
At $\mathrm{Pe}=0$, $p(x)$ is given by the Boltzmann distribution
\begin{equation}
    p(x)=\sqrt{\frac{k}{2\pi k_BT_c}} e^{-kx^2/2k_BT_c}\,.
\end{equation}
It stays approximately Gaussian for weak activity ($\mathrm{Pe}\lesssim 5$), while at higher activities ($\mathrm{Pe}\geq 5$), exponential tails appear, and only the center of the distribution remains Gaussian.
The Gaussian center of the distribution is due to the background thermal noise, as was also observed in the experiment~\cite{krishnamurthy2016micrometre}.
The tails arise from rare ``head-on collisions'' of persistent active particles with the probe, which result in large displacements.
In \textcolor{black}{a regime of intermediate $\mathrm{Pe}$ values where our numerical $p(x)$ exhibit a quadratic center and linear tails on the employed logarithmic scale}, our data can be fitted well by a superposition of Gaussian and Laplacian distributions \textcolor{black}{(fits not shown)}, as proposed in Refs.~\cite{leptos2009dynamics-of-enhanced-tracer-diffusion, thiffeault2015distribution}.
At the highest $\mathrm{Pe}$-values considered ($\mathrm{Pe}=10^2$), the tails are no longer purely exponential, and the fit fails.

Rescaling the distributions by their standard deviations $\sqrt{\sigma}$ makes them collapse when they are approximately Gaussian (so-called diffusive scaling~\cite{leptos2009dynamics-of-enhanced-tracer-diffusion,thiffeault2015distribution}).
Figure~\ref{fig:position-distr}(b) shows that increasing the bath activity $\mathrm{Pe}$ makes the tails of $p(x)$ more non-Gaussian.   
The lower two panels in Fig.~\ref{fig:position-distr}(c)-(d) show raw and scaled $p(x)$ for different probe sizes (color code) at the highest activity $\mathrm{Pe}=100$. 
As expected, the distributions are non-Gaussian for small relative probe diameters (blue curves) and gradually become Gaussian upon increasing $d_0$ (red curves).

To further quantify the conclusion that decreasing bath activity or increasing probe size makes the active noise more Gaussian, we investigated the variance and the kurtosis of the position distribution. (The odd moments like mean and skewness vanish due to symmetry, so that we focus on the even moments.)

\begin{figure}[t]
    \centering
    \includegraphics[width=\linewidth]{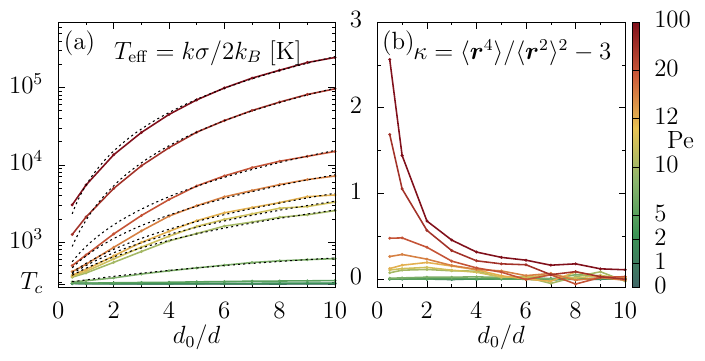}
    \caption{Second and fourth cumulants of the position distribution $p(x)$ over the relative probe size $d_0/d$ for different activities (color code).
    Dashed lines in (a) are quadratic fits.
    Data measured at $k_{\min}$ and $T_c$.}
    \label{fig:probe-size}
\end{figure}

We have already established in the comparison paper~\cite{partI}, that the position variance $\sigma\equiv\langle\bm r^2\rangle$ controls the average thermodynamic performance of the active heat engine, regardless of the shape of the position distribution. 
In Fig.~\ref{fig:probe-size}(a), $\sigma$ is represented in terms of the equivalent effective temperature $T_{\rm eff}= k\sigma/(2k_B)$~\cite{holubec2020active-brownian-heat-engines, partI}, which exhibits an approximately quadratic growth both in terms of probe size $d_0$ and activity $\mathrm{Pe}$, except for the passive result ($\mathrm{Pe}=0$) at the bottom, for which $T_{\text{eff}}\equiv T_c$. 
This behavior is in accord with the theoretical expectation, 
$T_{\rm eff} = T + v_a^2 d^2 / [2 \mu k_B(\mu k  + d D_r)]$,
valid for the reduced effective AOUP and ABP models~\cite{holubec2020active-brownian-heat-engines, partI}. 
In contrast, considering only the Gaussian center to define an effective temperature from simulations or experiments would generally underestimate the phenomenologically correct $T_{\rm eff} = k\sigma/(2k_B)$ and the associated work and heat~\cite{krishnamurthy2016micrometre}. 
As a technical remark, we note that for large activities and probe sizes, it is important to consider a bigger system with $N=500$ ABPs and the same density to avoid finite-size effects. 
We observed that $T_{\rm eff}$ is noticeably underestimated for $d_c\geq\SI{5}{\micro\metre}$ and $N=10^2$.

The excess kurtosis $\kappa$ in Fig.~\ref{fig:probe-size}(b) quantifies how much the distributions deviate from a Gaussian ($\kappa=0$). 
In agreement with our conclusions from Fig.~\ref{fig:position-distr}, we observe highly non-Gaussian effects for small probes $d_0\approx d$ and high activities ($\mathrm{Pe}\gtrsim 10$) when the measured kurtosis approaches $\kappa=3$, corresponding to that for the Laplacian distribution $p(x)= a e^{-a|x|}/2$.
An increase in probe size leads to a decline in $\kappa$, as the noise fluctuations eventually become Gaussian, in accord with the central limit theorem and literature results.
Similar leptokurtic distributions were theoretically predicted for displacement distributions in a 1D model with shot noise~\cite{roldan2024shot-noise-active-bath} and experimentally observed for tracer dynamics on cell substrates~\cite{grossmann2024non-gaussian-displacements}.
Gaussian force and position distributions were found for an underdamped model of a large, heavy probe in an active bath~\cite{schmid2022can-we-tell, schmid2024force-renormalization}.

\subsection{Active noise autocorrelations}
\label{sec:ACFs}
To identify the parameter regime where the active noise $\bm F$ in Eq.~\eqref{eq:eom-probe} is not only Gaussian but also exponentially correlated, as required for a perfect mapping between the complete many-body setup and the reduced single-particle AOUP model, we sampled its ACF in our simulations at constant stiffness $k=k_{\min}$ and temperature $T=T_c$, using a high sampling rate of $\mathrm{d}t = \SI{20}{\micro\second}$ instead of the experimental $\Delta t=\SI{2}{\milli\second}$.
We checked that the results are not visibly dependent on the trap stiffness, as it is expected since the active bath is coupled to the potential only indirectly and weakly, via the interactions with the probe particle.
When analyzing experimental data, one can approximately access the ACF of $\bm F$ by studying the autocorrelation of $\dot{\bm r} + \mu_0 k \bm r$, which is proportional to the sum of active and white noise $\mu_0 \bm F + \sqrt{2 D^0_t}\bm\xi$. 
In experiments, both quantities will be integrated over a sampling time interval $\Delta t$. 
Furthermore, since we found the cross-correlation $\langle\bm F(t)\cdot\bm\xi(0) \rangle$ to be negative in our simulations~\cite{partI}, we expect it to also affect the ACF measured in experiments. 
Accessing the active noise ACF in experiments can thus be challenging. Nevertheless, our simulation results should be sufficiently generic to provide quantitative estimates for parameter regimes where the active noise in experiments is exponentially correlated. Moreover, for the stiff traps employed in recent experiments~\cite{krishnamurthy2016micrometre}, the more relevant requirement for successfully modeling the experiments with an effective AOUP is the presence of Gaussian noise fluctuations, which is easier to verify than the condition of exponential noise ACFs.

\begin{figure}[t]
    \centering
    \includegraphics[width=\linewidth]{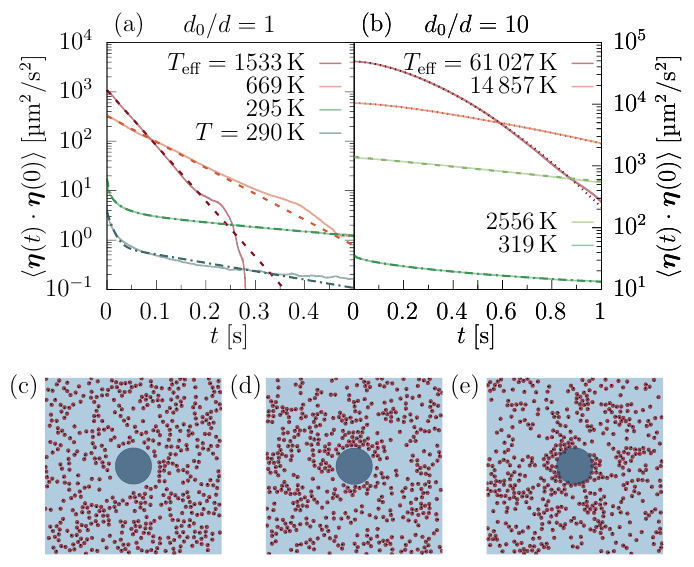}
    \caption{
    Noise autocorrelation functions (ACFs) for a passive tracer in an active bath composed of ABPs with a persistence time $D_r^{-1} = \SI{1}{\second}$.
    (a),(b) show ACFs of the active component $\bm\eta=\mu_0\bm F$ of the noise in Eq.~\eqref{eq:eom-probe}, as measured in the simulations, and (c)-(e) the corresponding snapshots of the active bath. Identical colors in (a) and (b) correspond to the same bath activity (swim speeds of ABPs) but different effective temperatures $T_{\rm eff}$ and probe sizes $d_0$.
    For $d_0=d$ in (a), the ACF can be fitted by a weighted sum~\eqref{eq:stretchExp} of an exponential and a stretched exponential with stretching exponent $1/2$ (dash-dotted lines), and at higher activities by an exponential (dashed lines), see appendix of~\cite{partI} for more data.
    For the larger probe, $d_0=10d$ in panel (b), dashed lines mark exponential fits at small and intermediate activities.
    For $\mathrm{Pe}\gtrsim 20$, the decay can be fitted by the compressed exponential in Eq.~\eqref{eq:comprExp} (dotted lines). The snapshots in the lower panels were taken for $d_0/d = 10$ and $T_{\rm eff}=\SI{319}{\kelvin}$ in (c), $T_{\rm eff}=\SI{2556}{\kelvin}$ in (d), and $T_{\rm eff}=\SI{61027}{\kelvin}$ in (e).
    }
    \label{fig:acfs-probe-size}
\end{figure}

Several example ACFs for the active noise obtained from our simulations are shown in Fig.~\ref{fig:acfs-probe-size}(a) for $d_0=d$.
The general overall trend is a speed-up of the decay with increasing activity.
At high activities ($\mathrm{Pe}\gtrsim 20$), the ACFs decay exponentially, indicated by the dashed exponential fits \textcolor{black}{(the `bumps' in ACFs at later times are artifacts caused by insufficient statistics of the simulated data and thus we aim to fit the initial parts only)}. 
The passive and low-activity ACFs exhibit a two-step decay, revealing some non-hydrodynamic short-time contributions.
They are better approximated by a weighted sum of an exponential and a stretched exponential (dash-dotted lines),
\begin{equation}
    c_1 e^{-\sqrt{t / \tau_1}} + c_2 e^{-t / \tau_2}\,,
		\label{eq:stretchExp}
\end{equation}
with $\tau_2/\tau_1\approx 10^2$, and $\tau_1\approx \SI{e-2}{\second}$.
The same form of the noise ACF has previously been observed in a similar system with a different interaction potential \cite{speck2023effective-dynamics}.

For the larger relative probe size $d_0/d = 10$ and high activities ($\mathrm{Pe}\gtrsim 20$) in Fig.~\ref{fig:acfs-probe-size}(b), we find the ACFs to decay more rapidly.
They can be fitted by compressed exponentials
\begin{equation}
    ce^{-(t/\tau)^\alpha}
    \label{eq:comprExp}
\end{equation}
with $\alpha>1$, shown by \textcolor{black}{orange ($\alpha=1.36$) and red ($\alpha=1.64$)} dotted lines in Fig.~\ref{fig:acfs-probe-size}(b).

We attribute these behaviors of the ACF to different spatial configurations and degrees of clustering of the active bath particles induced by the interactions with the probe~\cite{Jeanine2024}. 
For dilute active systems below densities at which global motility-induced phase separation occurs, the mean cluster size was found to grow linearly with the swim speed of the ABPs~\cite{theurkauff2012dynamic-clustering, speck2013dynamical-clustering}.
The simulation snapshots in Figs.~\ref{fig:acfs-probe-size}(c)-(e) and the distributions of local densities in Fig.~\ref{fig:local-density} in App.~\ref{app:local-density} indeed suggest an increased clustering of the bath ABPs around the probe, with growing activity.
In the passive and low-activity regimes ($\mathrm{Pe}\lesssim 20$), when the ACF has the form of a stretched exponential, the bath in Fig.~\ref{fig:acfs-probe-size}(c) appears fluid-like.
When the bath activity is increased further, within the activity regime exhibiting an exponential active noise ACF, the ABPs form small motility-induced clusters, mainly located close to the probe, which is indicative of heterogeneous nucleation and is illustrated in Fig.~\ref{fig:acfs-probe-size}(d). 
For the highest activity displayed in Fig.~\ref{fig:acfs-probe-size}(e), the dominant interaction of the probe appears to be with small active clusters, condensing onto the probe, which presumably induces the faster decay of the ACF. 

\section{Fluctuations}
\label{sec:fluctuations}
Let us now turn to the fluctuating thermodynamics of the active heat engine.
According to the stochastic energetics formalism established by Sekimoto~\cite{sekimoto1998langevin-thermodynamics, sekimoto2010stochastic-energetics}, the stochastic work and heat per cycle are obtained by splitting up the rate of change of the probe's potential energy according to $\dot{U}=\dot{k}\bm r^2/2 + k\bm r\cdot\dot{\bm r}$.
The contribution to $\dot{U}$ due to a variation of the externally controlled parameter $k$ represents the stochastic power flowing into the system. 
The work performed on the engine per cycle thus reads~\cite{partI}
\begin{equation}
    w = \frac{1}{2}\int_0^{t_p}\dot{k}(t) \bm r^2(t) \,\mathrm dt\,.
\end{equation}
The second contribution to $\dot{U}$ is identified as the rate of heating.
The stochastic heat input into the engine per cycle is then
\begin{equation}
   q_{\rm in} = \int_0^{t_p} k(t) \bm r(t)\cdot\dot{\bm r}(t) \Theta(\dot{\sigma})\,\mathrm dt\,.
\end{equation}
The Heaviside step function $\Theta(\dot{\sigma})$ is nonzero when the variance increases ($\dot{\sigma}>0$), signaling phases during which heat, on average, flows into the working medium. These are the branches II (literal heating) and III (\textcolor{black}{expansion} at maximum temperature and activity) of the cycle.
The ratio of output work $-w$ and input heat $q_{\text{in}}$ defines the stochastic efficiency, $\eta_s = -w/q_{\text{in}}$~\cite{verley2014efficiency-fluctuations,verley2014unlikely-carnot-efficiency}. 

\subsection{Full model with active particle bath}
\begin{figure}[t]
    \centering
    \includegraphics[width=\linewidth]{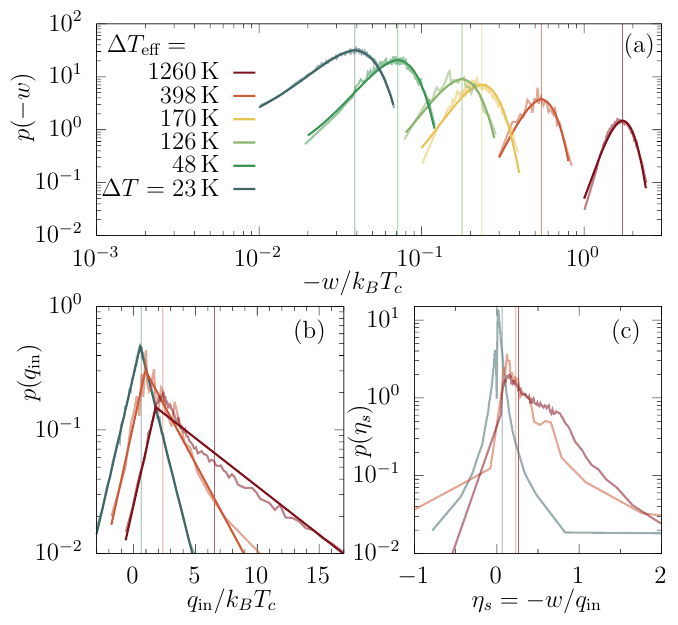}
    \caption{Distributions of (a) stochastic output work $-w$, (b) input heat $q_{\text{in}}$ and (c) efficiency $\eta_s$ for several $T_{\rm eff}$ (color code) and $d_0=d$,  obtained from simulations, and their mean values (vertical lines).
    For each color, the more opaque solid lines in (a) and (b) represent the model fits described in the text.
    The work distributions $p(-w)$ have a Gaussian center and exponential tails and are (linearly) symmetric with respect to their mean.
    The heat distributions $p(q_{\rm in})$ become increasingly asymmetric with growing $T_{\text{eff}}$ and can be approximated by exponentials with different decay rates on both sides of the maximum.
    (c) Distributions $p(\eta_s)$ of the stochastic efficiency.
    }
    \label{fig:work-distr}
\end{figure}
For notational conciseness, we will distinguish the probability densities by their dependent variable in the following, using the same symbol $p$ for all of them. 
The distributions $p(-w)$ of the stochastic work per cycle obtained from simulations of the active heat engine are shown as translucent lines in Fig.~\ref{fig:work-distr}(a). 
Vertical lines at the mean work, $W_{\rm out}=-\langle w\rangle$, intersect with the maxima of the distributions, showing $p(-w)$ are symmetric around $-\langle w\rangle$, which is a result of performing the cycle almost quasi-statically~\cite{holubec2021fluctuations-in-heat-enginges}. 
(That the distributions in the figure appear asymmetric is caused by the double logarithmic scale.)
The distributions exhibit a Gaussian center around the mean work and exponential tails, which is underscored by fitting Gaussian-Laplacian distributions (straight opaque solid lines) to the simulation data.
This shape of the work distribution is commonly reported in the literature~\cite{seifert2004distribution-of-work, engel2009asymptotics-of-work, jarzynski2011equalities, holubec2015asymptotic-behavior-of-work, holubec2021fluctuations-in-heat-enginges}. 
The Gaussian center corresponds to a large number of `typical trajectories' yielding output work close to the mean value and is a universal consequence of the central limit theorem. 
On the other hand, the tails correspond to rare independent collisions, yielding exceptionally large or small values of work.

The distributions $p(q_{\text{in}})$ of the stochastic heat accepted by the engine per cycle obtained from the simulations are shown as translucent lines in Fig.~\ref{fig:work-distr}(b), together with fits by asymmetric Laplacian distributions. 
The heat distributions have broad asymmetric tails around a Gaussian center, which is also much narrower than that of the work distributions.
This indicates that work done by cyclic heat engines is a self-averaging quantity, whose Gaussian character improves with increasing cycle duration, while the heat fluctuations can remain large even for quasi-static driving~\cite{Holubec2018HE,holubec2021fluctuations-in-heat-enginges}. \textcolor{black}{
The main source of this qualitative difference between the two stochastic processes is that stochastic heat corresponds to transitions between the system's energy levels and thus, in general, strongly depends on the system's initial and final states during the cycle, which control the increase in internal energy. Work, on the other hand, corresponds to the occupation times of the individual energy levels, and thus the corresponding dependence becomes negligible when the cycle duration is long compared to the system's relaxation time. For a detailed intuitive explanation of the differences between the two stochastic processes, see Refs.~\cite{Holubec2018HE, holubec2021fluctuations-in-heat-enginges}.}

Figure~\ref{fig:work-distr}(c) shows the distributions $p(\eta_s)$ of the stochastic efficiency from our simulations.
They have very heavy tails, and all of their moments diverge due to the sensitivity of the efficiency to small denominators~\cite{esposito2015efficiency-statistics-at-all-times}. 
Higher physical relevance than $p(\eta_s)$ has the large deviation function of the stochastic efficiency~\cite{proesmans2015stochastic-efficiency, verley2014unlikely-carnot-efficiency, martinez2015brownian-carnot-engine, verley2014efficiency-fluctuations, gingrich2014efficiency-and-large-deviations}.
We decided to calculate the stochastic efficiency distributions because their complex shape poses a stern test for the validity of the mapping to the AOUP model.
The efficiency distribution for zero activity is bimodal, with peaks at negative and positive $\eta_s$, corresponding to heat engine and refrigerator modes of operation~\cite{esposito2015efficiency-statistics-at-all-times}. 
With increasing activity, we find that $p(\eta_s)$ becomes unimodal with a maximum at a positive $\eta_s$.

\subsection{Effective AOUP model}
As our main results, we demonstrate in this section that numerical distributions of work, heat, and efficiency from the many-body system of Sec.~\ref{sec:model} can be reproduced by an effective one-particle AOUP model when the active noise $\bm F$ is Gaussian during the whole cycle and exponentially correlated during the large-activity part of the cycle.

The AOUP model is described by coupled overdamped Langevin equations for the position $\bm r$ of a probe that is actively driven, namely by the active noise term $\bm v$, which can be interpreted as the swim speed of an active particle,
\begin{align}
    \begin{split}
        \dot{\bm r} &= -\mu k \bm r + \bm v + \sqrt{2D_t}\bm\xi_t\,,\\
        \tau\dot{\bm v} &= -\bm v + \sqrt{2D_a}\bm\xi_a\,,
    \end{split}
    \label{eq:eom-aoup}
\end{align}
where the zero-mean, unit-variance Gaussian white noises $\bm\xi_t$ and $\bm\xi_a$ represent fluctuations in an equilibrium solvent~\cite{cates2021statistical-mechanics-AOUP,dabelow2021irreversibility-AOUP}.
The equivalent of the force exerted on the probe by the active bath in Eq.~\eqref{eq:eom-probe} is thus given by $\bm F = \bm v/\mu$, i.e.,~by \textcolor{black}{the Gaussian swim force of an effective microswimmer in Eq.~\eqref{eq:eom-aoup}.
The validity of this mapping has been recently proven for a probe particle in a granular gas~\cite{Caprini2024}, and it could hold more generally for thermodynamic interfaces in active baths~\cite{suchanek2023irreversible}.}
After formally solving Eq.~\eqref{eq:eom-aoup}, one finds 
the velocity autocorrelation function \textcolor{black}{
$\langle\bm v(t)\cdot\bm v(0)\rangle = 2D_a e^{-t/\tau} / \tau$}, where $\tau$ plays the role of a rotational diffusivity or persistence time of the effective microswimmer. 
The values of the mobility $\mu$, trap stiffness $k$, and diffusivity $D_t$ in the effective model~\eqref{eq:eom-aoup} remain those of Sec.~\ref{sec:model}. 
The active noise intensity $D_a$, and the persistence time $\tau$, are determined from fits to our above simulation data for the passive tracer in the active bath.

In Sec.~\ref{sec:coarse}, we found that the active noise ACF from the simulations is exponentially correlated for sufficiently high activity ($\mathrm{Pe}\gtrsim 10$) for small probes ($d_0\approx d$)~\cite{partI, speck2023effective-dynamics}, and intermediate activity ($10\lesssim\mathrm{Pe}\lesssim 20$) for larger probes ($d_0\gtrsim 5d$).
We also found that sufficiently large probes experience Gaussian active noise. 
For the parameters used in our study, the active noise was always close to Gaussian for $d_0/d \sim 10$.
To fit the AOUP model to our simulations, we thus extract the persistence time $\tau$ from the decay of the measured active noise ACF at the maximum P{\'e}clet number throughout the cycle. 
In the same spirit, the magnitudes of the ACFs could be used to determine also $D_a$, but this would require measuring the ACF for all P{\'e}clet numbers during the cycle. 
We therefore determined $D_a$ by instead requiring that the mean work per cycle in the effective model equals that found in the simulations. 
Therefor, we used the analytical expression for the mean work $W_{\rm out}$ done in the considered quasi-static cycle by a heat engine based on the effective AOUP~\cite{partI},
\begin{widetext}
    \begin{equation}
        W_{\rm out} = k_B(T_h-T_c)\log\frac{k_{\max}}{k_{\min}} + k_B(\tau\mu k_{\min} + 1) \left[ \Delta T_{\rm eff} - (T_h - T_c) \right] \log\frac{k_{\max}(\tau\mu k_{\min} + 1)}{k_{\min}(\tau\mu k_{\max} + 1)}\,,
    \end{equation}
\end{widetext}
where the effective temperature difference~\cite{holubec2020active-brownian-heat-engines,partI}
\begin{equation}
    \Delta T_{\rm eff} = T_h - T_c + \frac{D_a}{\mu k_B(\tau\mu k_{\min} + 1)}
\end{equation}
neglects the effect of a small swim speed during the cold isotherm ($T=T_c$) in the effective model, for simplicity~\cite{partI}.
The noise intensity of the effective model can then be expressed as
\begin{equation}
    \frac{D_a}{\mu} = \frac{W_{\text{out}} - k_B(T_h-T_c) \log\frac{k_{\max}}{k_{\min}}}{\log\frac{k_{\max}}{k_{\min}} - \log\frac{1+\tau\mu k_{\max}}{1+\tau\mu k_{\min}}}
    \label{eq:Da}
\end{equation}
where $W_{\rm out}$ is the average work numerically measured in the simulation.

Let us now test the ability of the effective AOUP model to predict the fluctuations of work, heat, and efficiency in the model bacterial heat engine.
The bath activity $\mathrm{Pe}$ in the considered protocol varies between a negligible minimum and a maximum value.
Our analysis in Sec.~\ref{sec:coarse} implies that low activity universally induces a double-exponential decay of the autocorrelation function (ACF). Thus, the actual active noise cannot be exponentially correlated throughout the entire cycle.
Nevertheless, the considered range of trap stiffnesses and mobilities implies that the relaxation time of the probe position in the harmonic trap, $(\mu_0 k)^{-1}$, is on the order of milliseconds, i.e., much shorter than the relaxation time of the active noise in a realistic parameter regime.
The stiff trap thus filters out correlations in the active noise occurring at timescales longer than $(\mu_0 k)^{-1}$ from the probe dynamics.
During the brief relaxation episodes in the trap, the noise ACF changes only very slowly and can thus approximately be fitted by an exponential even when it is globally non-exponential.

\begin{figure}
    \centering
    \includegraphics[width=\linewidth]{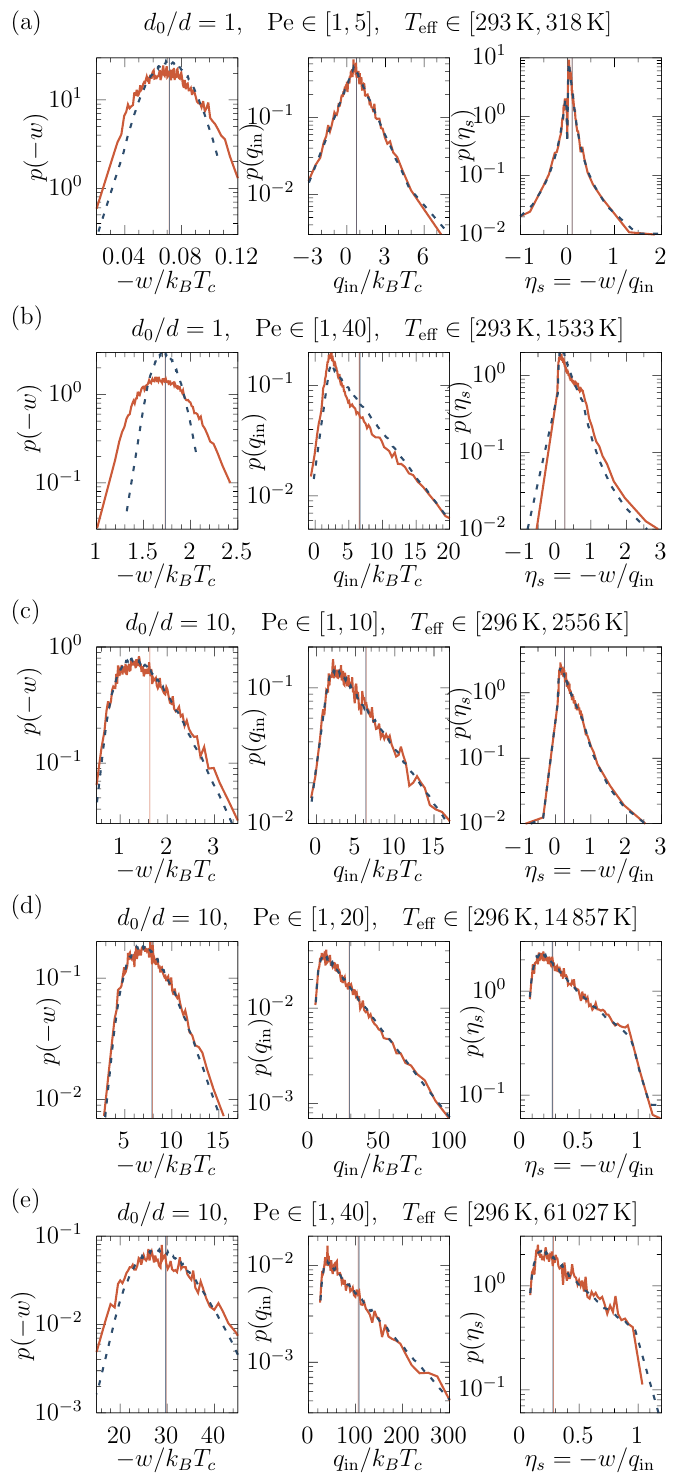}
    \caption{\color{black} Distributions of stochastic work $-w$, heat $q_{\rm in}$, and efficiency $\eta_s$ in the effective one-particle AOUP model (dashed lines) compared to the many-body system (solid lines).
    Each panel (a)-(e) represents one parameter set, consisting of the probe diameter $d_0$ and the range of bath activities $\mathrm{Pe}$, as indicated in their respective titles, together with the resulting range in effective temperature $T_{\rm eff}$.
    Vertical lines represent the mean values of the respective observables, which essentially coincide for both models~\cite{partI}.
    }
    \label{fig:AOUP}
\end{figure}

Indeed, Figs.~\ref{fig:AOUP}(c)-(e) corroborate that, for a large probe ($d_0 = 10d$), for which the active noise is Gaussian for all $\mathrm{Pe}$, the distributions obtained from the effective AOUP model and the simulations agree nicely, regardless of the precise form of the ACF, {\color{black} which is exponential only for Fig.~\ref{fig:AOUP}(c) but compressed exponential for Fig.~\ref{fig:AOUP}(d),(e), cf. Fig.~\ref{fig:acfs-probe-size}(b)}.
This holds, in particular, for the broad heat and efficiency distributions. 
Deviations in the tails of the work distributions are minimal in Fig.~\ref{fig:AOUP}(d) and increase as the part of the cycle during which the active noise ACF is non-exponential grows.
Similarly, the probability densities nicely agree for a small probe ($d_0=d$) and low activity in Fig.~\ref{fig:AOUP}(a), where the noise is approximately Gaussian with $\kappa\lesssim 0.5$, {\color{black} although the ACF decays as a stretched exponential, cf. Fig.~\ref{fig:acfs-probe-size}(a)}.
\textcolor{black}{Increasing bath activity and thus non-Gaussianity of the corresponding noise} render the agreement worse, as shown in Fig.~\ref{fig:AOUP}(b), for which the ACF is exponential, but the position distribution is highly non-Gaussian.

To close this section, we note that increasing the probe diameter increases the relaxation time of its position in the harmonic trap. 
This means that for $d_0/d = 10$, as used in Figs.~\ref{fig:AOUP}(c)-(e), the cycle is perceived as effectively 10 times faster by the probe than for $d_0/d = 1$, as in Figs.~\ref{fig:AOUP}(a),(b). 
As a result, the cycles for $d_0/d = 10$ are no longer approximately quasi-static.
Instead of symmetric work distributions~\cite{seifert2004distribution-of-work}, they, therefore, generate a long exponential tail of positive work values.

\newpage
\newpage

\section{Conclusion and Outlook}
\label{sec:conclusion}

We studied a numerical model of a cyclic bacterial heat engine for varying sizes of the colloidal tracer and different swim speeds of the surrounding active bath particles.
We identified parameter regimes where an effective one-particle AOUP toy model exhibits the same stochastic thermodynamics as a full-fledged Brownian dynamics simulation of the engine.
The most important condition is that active noise experienced by the tracer is Gaussian, which occurs for moderate bath activities when the tracer particle is several times larger than the bath particles  and at low activities even for comparable particle sizes. 
Secondly, the noise autocorrelation function should be exponential, which occurs for intermediate activities. 
However, the latter requirement can be softened by using stiff traps, inducing fast relaxation of the tracer position, and filtering out the non-exponential part of the noise autocorrelation function.
Such stiff traps are commonly used in current experiments with bacterial heat engines~\cite{krishnamurthy2016micrometre}, making the AOUP an ideal model to reproduce or analyze the measured data.
Noteworthily, we have performed an equivalent analysis using the active Brownian particle (ABP) model~\cite{holubec2020active-brownian-heat-engines}, which is very similar to the AOUP except that the active noise fluctuations are non-Gaussian, finding no nontrivial parameter regime where the fluctuations of the full-fledged model and the ABP model would agree.

When the mapping of the tracer dynamics in the (many-body) active bath to that of the effective active Ornstein-Uhlenbeck particle works, the latter can be used to predict, with good to perfect accuracy, work, heat, and efficiency distributions or the corresponding large deviation functions, which are usually hard to measure in experiments. 
Besides computer simulations, one can solve for these distributions even analytically for simple piecewise constant protocols, using the techniques described in Refs.~\cite{Holubec2018HE,holubec2021fluctuations-in-heat-enginges}. To be specific, our analysis shows that the effective models should capture the experimental results in Ref.~\cite{krishnamurthy2016micrometre} for tracers (slightly) larger than the bacteria and for bacterial swim speeds $v_a\simeq\mathcal O(\SI{10}{\micro\metre\per\second})$. Our operational approach to coarse-graining could also be useful to find effective models in other situations.

\section*{Acknowledgements}
V.H. acknowledges the support of Charles University through project PRIMUS/22/SCI/009. K.K. acknowledges financial support from the Deutsche Forschungsgemeinschaft for computational resources.

\appendix

\section{Local particle-density distributions in the active bath}
\label{app:local-density}
\begin{figure}
    \centering
    \includegraphics[width=\linewidth]{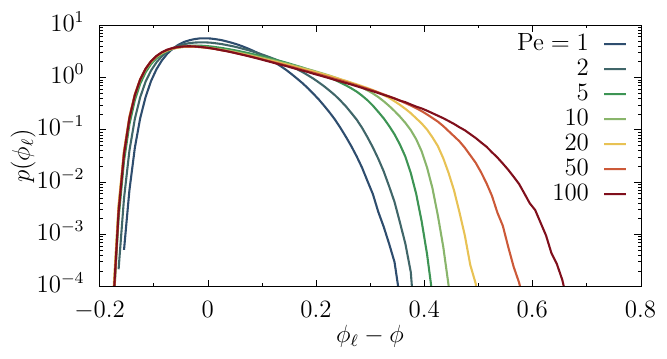}
    \caption{The distribution $p(\phi_{\ell})$ of the local bath-particle density $\phi_{\ell}$ around the global packing fraction $\phi=0.2$ is unimodal, but broadens with increasing activity.
    This indicates the growing size and number of dense motility-induced particle clusters.}
    \label{fig:local-density}
\end{figure}

To quantify the tendency of bath particles to cluster upon increasing their activity, as seen in Figs.~\ref{fig:acfs-probe-size}(d), (e), where the clustering is additionally fostered by the presence of the probe, we analyzed the distribution of local densities in the ABP bath also without the probe particle and for an increased particle number of $N=10^4$.
We used a Voronoi tessellation~\cite{rycroft2009voro++} to assign each particle $i$ an area $A_i$, defined by its Voronoi cell.
And we defined a local bath-particle density as the area fraction $\phi_{\ell, i} = \pi d^2/4A_i$ of bath ABPs of diameter $d$.
Computing a histogram for $\phi_{\ell,i}$ and averaging over $4000$ particle configurations then gave the empirical distributions $p(\phi_{\ell})$ shown in Fig.~\ref{fig:local-density}.
We observe that the tails representing regions of enhanced local density $\phi_{\ell}>\phi$ are becoming increasingly pronounced for high activities, suggesting a growing tendency of the ABPs to cluster.

\bibliographystyle{apsrev4-2}
\bibliography{refs}
\end{document}